# Generation of two-dimensional plasmonic bottle beams

Patrice Genevet [1,*], Jean Dellinger [2,*], Romain Blanchard [1], Alan She[1], Marlene Petit[2], Benoit Cluzel [2], Mikhail A. Kats[1], Frederique de Fornel [2] and Federico Capasso[1,&]

*these two authors have contributed equally to this work
[&] capasso@seas.harvard.edu
[1] School of Engineering and Applied Sciences, Harvard University, Cambridge, Massachusetts 02138, USA
[2] Laboratoire Interdisciplinaire Carnot de Bourgogne, UMR CNRS 6303, 9 avenue Savary, Dijon, 21078, France

**Abstract:** By analogy to the three dimensional optical bottle beam, we introduce the plasmonic bottle beam: a two dimensional surface wave which features a lattice of plasmonic bottles, i.e. alternating regions of bright focii surrounded by low intensities. The two-dimensional bottle beam is created by the interference of a non-diffracting beam, a cosine-Gaussian beam, and a plane wave, thus giving rise to a non-diffracting complex intensity distribution. By controlling the propagation constant of the cosine-Gauss beam, the size and number of plasmonic bottles can be engineered. The two dimensional lattice of hot spots formed by this new plasmonic wave could have applications in plasmonic trapping.

**I.Introduction**

Since their recent discovery, optical bottle beams [1-4]  have found applications ranging from dynamical trapping of atoms to sorting of particles in colloidal suspensions [5-9]. The interest of bottle beams for optical trapping arises from the oscillating intensity distribution, creating dark (or bright) focii surrounded in all directions by intense (or low) electromagnetic fields. This distribution of light, in the shape of a bottle, creates suitable intensity gradients to trap low (or high) index particles inside the dark (or bright) spots.

The emerging field of plasmonic trapping employs nano-structured metallic resonators [10-13] or propagating surface waves [14] to control the position of particles in the proximity of metallic structures. Plasmonic devices with reconfigurable 2-D grids of near-fields could be integrated into microfluidic systems, which is of interest for applications for example in biology [15].

In this letter, we experimentally demonstrate the two dimensional analogue of optical bottle beams. These surface waves, which present oscillating low and high intensity focii, are generated at the interface between a metal and a dielectric. In the dielectric half-space, the component of the surface plasmon electric field normal to the metal-dielectric interface obeys the Helmholtz equation:

$$\nabla^2 E_z + \varepsilon_d k_0^2 E_z = 0 \quad (1)$$

The surface plasmon propagation constant $k_{sp}$ satisfies the equation

$$k_{sp}^2 = k_x^2 + k_y^2 = \alpha^2 + \varepsilon_d k_0^2 = k_0^2 \frac{\varepsilon_d \varepsilon_m}{\varepsilon_d + \varepsilon_m} \qquad (2)$$

$k_0$ is the free space propagation constant $-\alpha$ is the inverse of the decay length length in the dielectric, $\varepsilon_d$ and $\varepsilon_m$ are the permittivity of diellectric and the metal, respectively. In recent work, we demonstrated the cosine-Gauss beam (CGB), a non-diffracting solution of Eqn. (1) [16]. The field profile of the CGB is given by:

$$E^{CGB}_z = A.f(x).\exp(jk_x x)\cos(k_y y)\exp\left(-\frac{y^2}{w_0^2}\right)\exp(\alpha z) \qquad (3)$$

where $w_0$ is the Gaussian waist and f(x) is described in [16]. Compared to a plasmonic Airy beam, another non-diffracting surface wave [17-20], CGB is a conical solution which can be viewed as the 2D projection of a free-space Bessel beam with finite spatial extent. As previously discussed in Refs. [16,21-23], CGBs are formed by the interference of two two-dimensional plane waves with intersecting directions of propagation. The straight lines along which constructive interference occurs define the propagation direction of the CGB, see Fig.1(a). Its propagation constant is given by $k_x^{CGB} = k_{sp}\cos\theta$, where $\theta$ is the half angle between the direction of propagation of the two plane waves components. By multiplying the two plasmonic plane waves by a Gaussian envelope of finite size, we demonstrated that the CGB remains non-diffracting in the paraxial approximation, while carrying finite amount of energy. While this envelope localizes the solution in a similar way as for Bessel Gauss beams [24], it also introduces negligible diffraction effects which are unnoticed due to the relatively short SPP the propagation distance, i.e. the main narrow lobe of the CGB diffracts very slowly as a beam with the same transverse dimension as the Gaussian envelope. Note that this envelope is significantly larger than the narrow width of the main intensity lobe of the CGB, explaining its non-diffracting behavior. The CGB propagates in a straight line with a constant and controllable phase velocity, which can be modified by changing the half angle $\theta$ between the two plane waves.

## II. Theory and description

In the present work, we use the interference between co-propagating non-diffracting surface waves of different propagation constants to create an array of plasmonic hot spots. One way to achieve such plasmonic intensity grid could consist in superposing two different CGBs (Fig1(b)). Instead here, we superimpose a plasmonic plane wave propagating collinearly with the CGB (Fig. 1(c)). This scheme not only generates the characteristic intensity modulation of plasmonic bottle beams but is also more suitable for experimental implementation of plasmonic trapping. In this specific case, the plane wave coupler can be arranged such that it will not spatially overlap with the CGB couplers. This allows us to dynamically control the position of the spots of intensity along the propagation direction (x): by adjusting the relative phase between the plane wave and the CGB, we can adjust the x-position of the trapping sites. The electric field distribution normal to the plane of our two-dimensional bottle beam is given by:

$$E_z = \left[ A.f(x).\exp(jk_x^{CGB}x)\cos(k_y^{CGB}y) + B\exp(jk_{sp}x) \right] \exp\left(-\frac{y^2}{w_0^2}\right) \exp(\alpha z) \quad (4)$$

Note that the last two terms, affecting both the CGB and the plane wave, account for the Gaussian envelope and for the characteristic exponential decay of SPPs in the dielectric for z>0. The intensity distribution of this bottle beam is plotted in Fig.1(c).

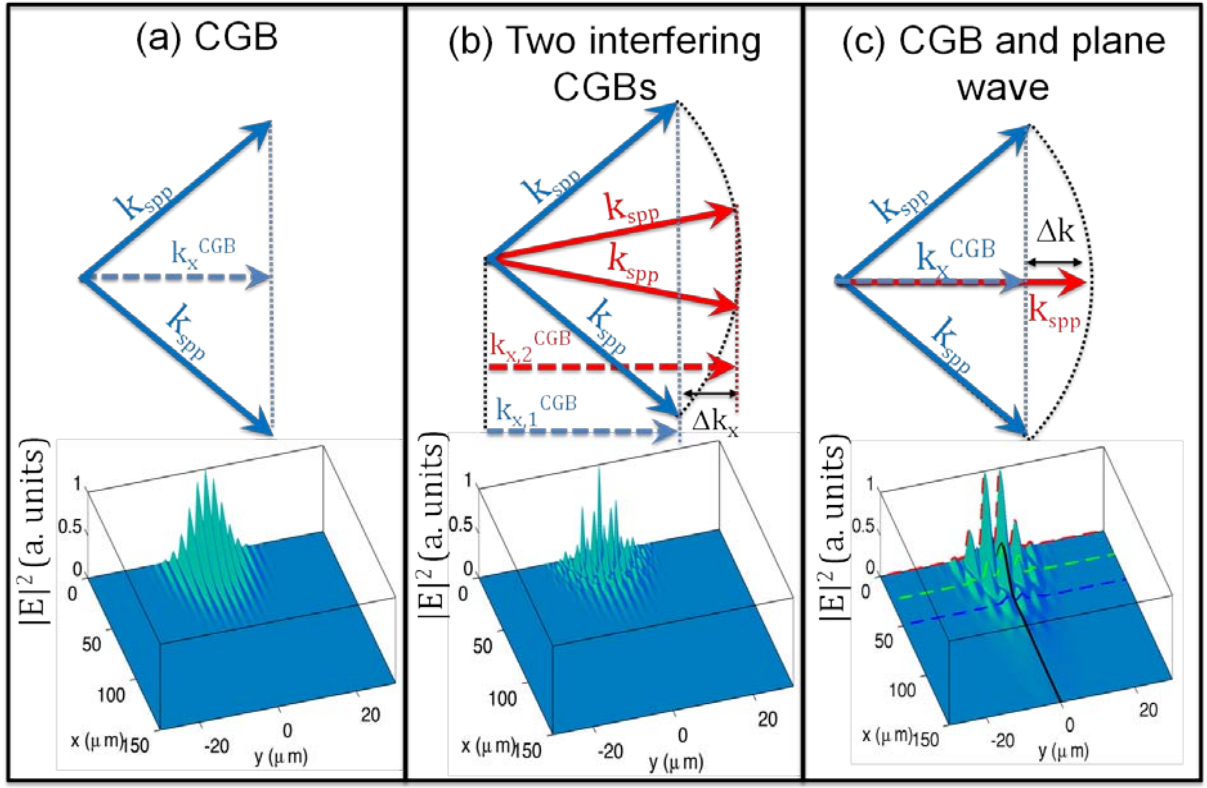

**Figure 1.** (a) The Cosine Gauss Beam (CGB) is formed by interfering two plane waves propagating at an angle. Its wavevector, $k_x^{CGB}$, is therefore strictly smaller than the plane waves wavevector $k_{spp}$. Two different methods to generate plasmonic bottle beams: (b) due to the wavevector mismatch between different CGBs ($\theta_1 = 10°$ and $\theta_2 = 20°$) interference effects will produce an in-plane intensity modulation along the propagation direction; (c) the same type of wavevector mismatch occurs when a CGB ($\theta = 10°$) is superposed onto a co-propagating plane wave. The bottom figures show the near-field intensity distribution (electric field component normal to the plane) obtained from analytical calculations. In (b) the amplitudes of the two CGBs are equal. In (c), since these two surface waves have different propagation constant, and therefore different propagation losses, we control the amplitude of the electric field of the plane wave (term B in Eqn. 3, scaled down to 1/1400 of that of the CGB, i.e. B=A/1400) to maximize the number of bottles. The parameters used in the calculations are: $k_{spp} = 8.7.10^6 + i\,1.7.10^4\,m^{-1}$, corresponding to SPPs excited by free space wavelength of about 730 nm [25] and $W_0 = 8.10^{-6}$. The dashed colored curves and the black curve locate the position of the intensity cross sections plotted in Fig.2.

The number of bottles along the propagation direction, N, is proportional to the decay length of SPP multiplied by the spatial frequency of the interference pattern, the latter being the wavevector mismatch between the CGB and the plane wave. The wavevectors of the CGB and the plane wave are respectively $k_x^{CGB} = \text{k}_{\text{spp}}.\cos\theta$ and $\text{k}_{\text{plane}} = \text{k}_{\text{spp}}$. The wavevector mismatch is expressed as $\Delta k_x = \text{k}_{\text{plane}} - k_x^{CGB} = \text{k}_{\text{spp}}.(1-\cos\theta)$, see illustration in Fig. 1(c). This gives $N \propto \frac{Re\{\text{k}_{\text{spp}}\}}{Im\{k_x^{CGB}\}}(1-\cos\theta)$. As the angle between the two components of the CGB increases, the bottles become more numerous and their size decreases. Fig. 2 (a) summarizes the evolution of both the length and the width of the bottle as a function of $\theta$ and shows the longitudinal and transverse field distributions for $\theta = 10^0$ (fig 2 b and 2c, respectively).

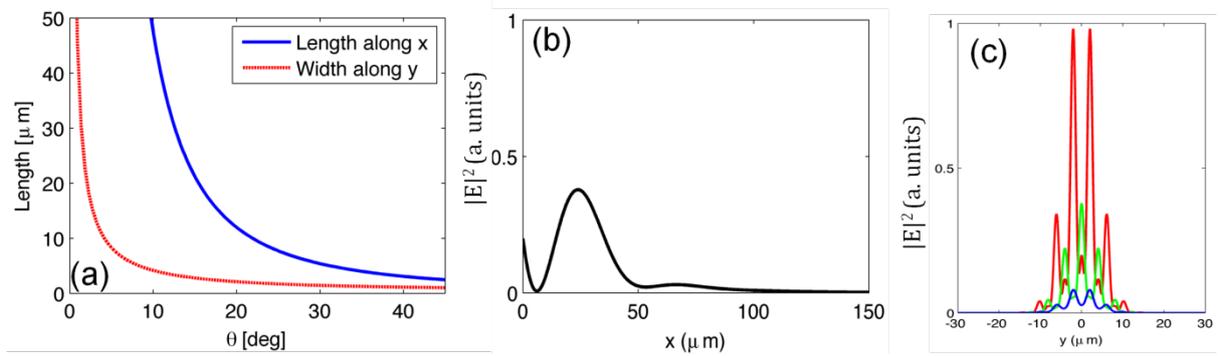

**Figure 2.** (a) Theoretical calculation of the length and width of the near-field bottles with respect to the half angle $\theta$ between the two plane waves creating the Cosine-Gauss beam. The length and width are calculated by considering the distance between two consecutive minima along x and y respectively. (b) shows the normalized near-field intensity along the propagation axis x, for y=0, as illustrated by the black curve in Fig.1(c). The red, green and blue curves in (c) indicate intensity distribution after several propagation distances x (respectively x = 0, 25, 50 µm) (also indicated in Fig.1(c) by the dashed lines normal to the propagation direction). The parameters used in the calculations are the same as those used in Fig1.

Due to the non-diffracting character of this solution, the interference pattern sets up a lattice of bottles with constant size and spacing (Fig. 1(c) and Fig. 2(a)). By changing the relative phase between the CGB and the plane wave, we can move the peaks in intensity back and forth along the x-direction (see movie in additional materials). Note that by changing the relative phase at a controlled rate, it is possible to impart precise momentum to trapped particles.

### III. Experimental results and discussions

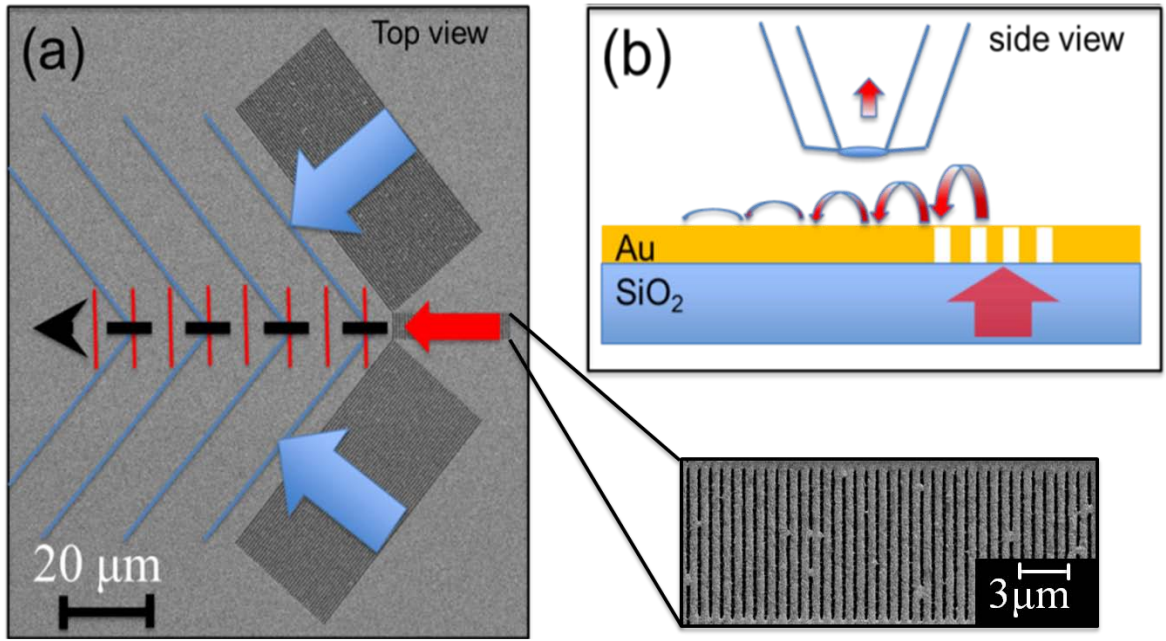

**Figure 3.** (a) Scanning electron micrograph of the plasmonic bottle beam coupler. The beam is formed by illuminating three gratings etched on the surface of a gold film. We designed the period of these gratings (fixed at 710nm) to match the wavelength of SPPs, in order to resonantly excite SPPs with illumination at normal incidence. Two of the gratings offset from each other by an angle, launch plane wave surface plasmons (blue lines) which interfere with each other along the symmetry axis to form the propagating Cosine-Gauss beam. An additional grating is then used to excite an on-axis surface plasmon plane wave (red lines), which in turn interferes with the cosine Gauss beam. The resulting interference pattern results in a periodic modulation of the intensity along the propagation direction,the bottle beam, as illustrated by the dashed black arrow. (b) Sketch of the near-field scanning optical microscope (NSOM) experimental configuration. The light is focused from the glass substrate side onto the gratings. The gratings launch the surface waves and the two dimensional near-field distribution is collected using the NSOM [26].

We used a focused ion beam (Zeiss NVision 40) to mill the plasmonic bottle beam couplers into a 150nm-thick gold layer. Prior to FIB milling, the gold film was template-stripped [27] from a silicon wafer onto a glass substrate in order to decrease surface roughness, thus decreasing scattering losses and increasing the SPP propagation distance. The two-dimensional SPP field distribution is acquired by collecting the in-plane field components using a near-field scanning optical microscope (NSOM) working in aperture mode [aperture].The structures are excited by a Ti:Sapphire laser collimated on the sample at normal incidence. The results presented in this letter have been obtained by tuning the emission and the detection to a specific wavelength of 735nm, i.e. at the wavelength which maximizes the coupling efficiency of the gratings. The NSOM images obtained for half angle of 10, 20 and 30 degrees are shown in the left panel of Fig.4.

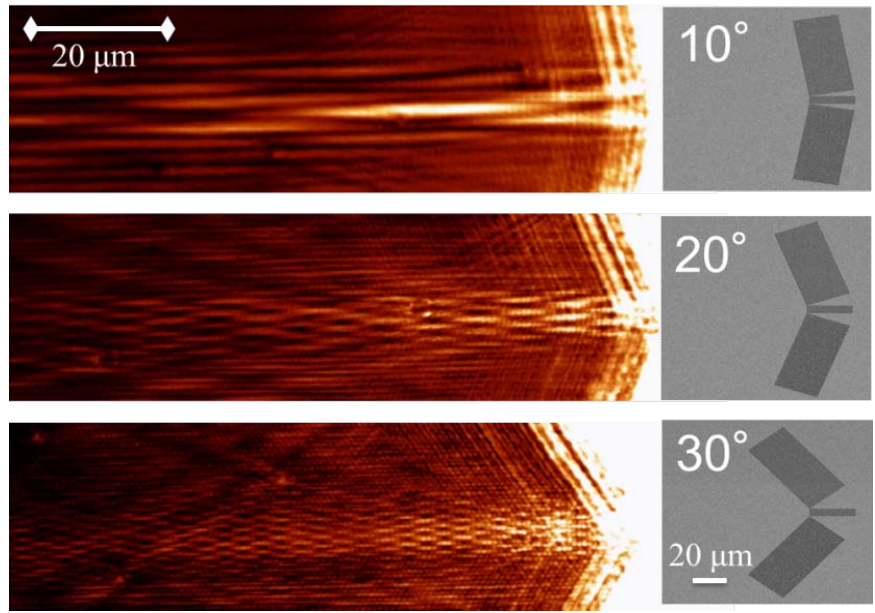

**Figure 4.** Near-field scanning optical microscope (NSOM) images showing the in-plane intensity of several plasmonic bottle beams. By increasing the angle between the CGB couplers, we can excite bottle beams with different beating periods. Scanning electron micrograph of the corresponding grating couplers are also presented on the right part of the NSOM images. The bottle waves are created on the leftmost part of the devices, i.e. where all the waves overlap. From the NSOM images, we can also appreciate the non-diffracting character of the beams which preserve the size and the shape of the intensity pattern throughout the propagation distance (estimated at 30μm for an air/Au interface for an incident wavelength $\lambda_0$ =735 nm).

In conclusion, we have demonstrated a new two-dimensional beam, the plasmonic bottle beam. This wave is generated by superimposing a non-diffracting cosine-Gauss beam (CGB), characterized by a small on-axis propagation constant, and a quasi-plane wave. The plasmonic bottle beam features a grid of intensity peaks and valleys which remains homogeneous due to its non-diffracting character. Interferometric experiments have been realized with other recently-demonstrated non-diffracting solutions such as Airy beams [28], but this type of periodic field distribution cannot be achieved due to their self-accelerating nature. The plasmonic bottle beam will find applications in plasmonic optical trapping and in other fields of research where an intensity lattice is required. Plasmonic bottles could be used to optically sort particles by trapping those with a specific size. Since the location of the plasmonic bottles is defined by the relative phase between the CGB and the plane waves, it is possible to trap particles at intensity spots while shifting the phase of the plane wave to create plasmonic tractor bottle beams, i.e. plasmonic fields that can transport trapped particles even against the plasmon stream [29,30]. By taking advantage of the cyclic nature of phase, it may be possible to build an optical ratchet, in which particles are continuously pulled or pushed along the beam. By arranging two or more bottle-beams around a central "working area", it might be possible to achieve full 2-D control on the position of particles trapped close to a metallic surface.

**Acknowledgments**

The authors acknowledge support from the National Science Foundation (NSF). The devices fabrication was performed at the Harvard Center for Nanoscale Systems (CNS) which is a member of the National Nanotechnology Infrastructure Network (NNIN). This research is supported in part by the Air Force Office of Scientific Research under grant number FA9550-12-1-0289. M. Kats is supported by the NSF through the Graduate Research Fellowship Program. B. Cluzel and F. de Fornel acknowledge the burgundy regional council for financial support through the PHOTCOM project.